\documentclass[aps,pra,twocolumn,showpacs,groupedaddress]{revtex4-1}
\usepackage{graphicx}
\usepackage{amsmath}
\usepackage{color}
\usepackage{notes2bib}
\usepackage{hyperref}

\begin{document}

\title{Addressing electron-hole correlation in core excitations of solids: An all-electron many-body approach from first principles}


\author{Christian Vorwerk}
\email[]{vorwerk@physik.hu-berlin.de}
\affiliation{Institut f\"ur Physik and IRIS Adlershof, Humboldt-Universit\"at zu Berlin, Berlin, Germany}
\affiliation{European Theoretical Spectroscopic Facility (ETSF)}
\author{Caterina \surname{Cocchi}}
\affiliation{Institut f\"ur Physik and IRIS Adlershof, Humboldt-Universit\"at zu Berlin, Berlin, Germany}
\affiliation{European Theoretical Spectroscopic Facility (ETSF)}
\author{Claudia \surname{Draxl}}
\affiliation{Institut f\"ur Physik and IRIS Adlershof, Humboldt-Universit\"at zu Berlin, Berlin, Germany}
\affiliation{European Theoretical Spectroscopic Facility (ETSF)}


\date{\today}

\begin{abstract}
We present an \textit{ab initio} study of core excitations of solid-state materials focusing on the role of electron-hole correlation.
In the framework of an all-electron implementation of many-body perturbation theory into the \texttt{exciting} code, we investigate three different absorption edges of three materials, spanning a broad energy window, with transition energies between a few hundred to thousands of eV.
Specifically, we consider excitations from the Ti $K$ edge in rutile and anatase $\textrm{TiO}_2$, from the Pb $M_4$ edge in $\textrm{PbI}_2$, and from the Ca $L_{2,3}$ edge in CaO.
We show that the electron-hole attraction rules x-ray absorption for deep core states, when local fields play a minor role.
On the other hand, the local-field effects introduced by the exchange interaction between the excited electron and the hole dominate excitation processes from shallower core levels, separated by a spin-orbit splitting of a few eV.
Our approach yields absorption spectra in good agreement with available experimental data, and allows for an in-depth analysis of the results, revealing the electronic contributions to the excitations, as well as their spatial distribution.
\end{abstract}


\maketitle

\section{Introduction}

An accurate first-principles description of excited-state properties in crystalline materials requires a high-level treatment of electron correlation.
Green's function-based many-body perturbation theory (MBPT) \cite{hedi65pr}, including the $GW$ approximation \cite{hybe-loui84prb,arya-gunn98rpp} and the solution of the Bethe-Salpeter equation (BSE) \cite{hank-sham80prb,Strinati}, represents the state of the art for calculating charged and neutral excitations in the optical region, respectively \cite{albr+98prl,bene-shir99prb,BSE:Rohlfing}. While the former yields quasi-particle energies, accounting for electron-electron correlation, the latter gives excitation energies of the electron-hole pairs. These methods are routinely applied on top of a mean-field description of the ground state obtained from density-functional theory (DFT) \cite{hohe-kohn64pr,kohn-sham65pr}.

Correlation effects enter the description of core excitations differently. Several approaches are adopted, often with good results, depending on the range of the transition energies \cite{XAS:Rehr, XAS:Tanaka}. Excitations from deep-lying core states, which are highly localized and decoupled from the surrounding electron cloud, are dominated by the long-range electron-hole ($e$-$h$) attraction. As such, these excitations have been successfully described by a constrained DFT-based scheme, the so-called \textit{core-hole} or the \textit{final-state rule} approximation \cite{FSRvsBSE}.
With the core-hole potential directly included in the calculation of the electronic states, this method is well suited for extended systems \cite{CoreHole_C3N4,CoreHole_BN,CoreHole_LiO2}, although requiring \textit{supercell} calculations to avoid spurious interactions between core holes of neighboring unit cells.
For shallower core states, however, the overlap between the initial core state and the conduction bands is usually more relevant and the local treatment of the electron-hole correlation becomes problematic  \cite{Olovsson2}. Indeed, the core-hole approximation is known to fail in case of excitations involving shallow core or even semi-core states, that may interact with the surrounding electrons \cite{Olovsson}.
Other mean-field methodologies, such as multiple scattering calculations \cite{XAS:Rehr, XAS:Rehr2, XAS:Rehr3}, are often successfully adopted for finite systems, but are hardly applicable to solids.

BSE-based approaches to core excitations have proven to overcome these issues and to be reliable for many crystalline materials.
However, most of the schemes proposed so far are either based on pseudo-potential approximation \cite{ocean,oceanII}, or, in an all-electron framework, on the description of the core states in terms of local orbitals as part of the valence region, neglecting spin-orbit coupling.
While the latter choice turned out to be effective in many cases \cite{Olovsson,Olovsson3,olov+13jpcm,shie+13pnas,Cocchi2}, this approximate treatment of core states is limited to a specific energy window and initial states. In particular, excitations from core levels with a non-negligible spin-orbit splitting need a relativistic treatment, even for low transition energies.
Excitations from $2p_{1/2}$ and $2p_{3/2}$ states in transition-metal oxides are an exemplary case that requires an enhanced spinorial treatment of the core electrons \cite{Blaha}.
In such systems local-fields effects are crucial for the spectra, and an accurate description of the $e$-$h$ exchange interaction is essential to achieve reliable results.

In this work, we aim at addressing correlation effects in core excitations of solids, by exploring the spectra of different edges of several materials, spanning a broad range of transition energies.
To do so, we adopt an \textit{ab initio} approach based on the solution of the BSE in an all-electron framework.
After presenting the formalism and the essence of its implementation into the \texttt{exciting} code \cite{exciting}, we show x-ray absorption spectra of three materials, focusing on excitations from the Ti $K$ edge in rutile and anatase $\textrm{TiO}_2$, from the $M_4$ edge in lead iodide, and from the Ca $L_{2,3}$ edge in CaO.
We discuss the role of the $e$-$h$ attraction and exchange, and analyze our results to gain insight into the nature of these excitations.

 \section{Methodology}
 \subsection{Theoretical background}
In many-body perturbation theory, absorption spectra are usually computed in three steps: First, the electronic structure of the system is calculated within DFT. On top of this, the quasiparticle (QP) energies and wavefunctions are determined to account for electron-electron interactions.  While calculations of the self-energy for valence and conduction states are nowadays routinely performed in the $GW$ scheme, quasi-particle corrections to the core region are, at present, still largely unexplored. Early attempts to estimate the self-energy of core states date back to the pioneering works of F. Bechstedt \cite{Bechstedt82} and L. Hedin \cite{Hedin2}. More recently, a numerical study on shallow $d$-core states has disclosed the challenges related to the QP correction further from the valence region \cite{Fuchs}. One of the main reasons is the poor starting point provided by DFT for such localized and dispersionles states, when semi-local exchange-correlation functionals are employed. Some degree of self-consistency in $GW$ or vertex corrections have to be included to overcome this limitation. While these methods are becoming the state of the art for evaluating the electronic properties of solids in the valence and conduction region \cite{scGW,vertex}, in theoretical core spectroscopy it is still common practice to mimic the QP correction of both unoccupied and core states with a scissors shift \cite{Blaha, Vinson}. The exact value of the corresponding operator is chosen such that the onsets of calculated and experimental spectra are aligned \cite{Blaha, Vinson}. Shifts of a few tens of eV are typically applied, with the contribution coming from the core states being largely dominating.
 
As a third step, neutral excitations are described by the \textit{ macroscopic response function}  $\bar{P}(\mathbf{r}_1, t_1,\mathbf{r}_1', t_1';\mathbf{r}_2, t_2,\mathbf{r}_2', t_2')$. This quantity is used to obtain the macroscopic dielectric function $\epsilon_M$ without the inversion of the microscopic dielectric function $\epsilon_{\mathbf{G}\mathbf{G}'}$ \cite{BSE:Onida, Bechstedt2015}. It is determined by a Dyson-like equation $\bar{P}=P_0+P_0\bar{\Xi} \bar{P}$, which connects the response function $\bar{P}$ with its non-interacting irreducible polarization function $P_0$ through the interaction kernel $\bar{\Xi}$. This is the so-called the Bethe-Salpeter equation \cite{Strinati}. As detailed in Ref. \cite{BSE:Rohlfing}, the matrix inversion $\bar{P}=[1-P_0\bar{\Xi}]^{-1}P_0$ is typically mapped onto an effective eigenvalue problem
 \begin{equation}\label{eq5_18}
   \sum_{c'u'\mathbf{k}'}H^{BSE}_{cu\mathbf{k}, c'u'\mathbf{k}'}A^{\lambda}_{c'u'\mathbf{k}'}=E^{\lambda}A^{\lambda}_{cu\mathbf{k}},
 \end{equation}
 where, in our case, the summation is limited to the initial core states, labeled by $c$, and the final unoccupied states, labeled by $u$. The core states are defined by the absorption edge considered, while the number of unoccupied states needs to be carefully converged.
 The eigenvalues $E^{\lambda}$ and the corresponding eigenstates $A^{\lambda}_{cu\mathbf{k}}$ uniquely define the two-particle excitations of the system.
 The former represent the excitation energies, and are typically associated with the binding energies $E_b$, which are usually defined as the difference between $E^{\lambda}$ and the energy of the lowest transition within the independent-quasiparticle approximation (IQPA), $E_b=E^{\lambda}-E_0^{IQPA}$. In this work, we adopt a scissors operator $\Delta$. Thus, the QP energy difference between core and conduction states reads as follows: $E^{QP}_{u\mathbf{k}}-E^{QP}_{c} \approx \epsilon_{u\mathbf{k}}-\epsilon_{c}+\Delta$, where $\epsilon_{u \mathbf{k}}$ and $\epsilon_{c}$ are the Kohn-Sham eigenvalues of the conduction and core state respectively. To avoid confusion with results obtained from $GW$ calculations, we chose the nomenclature \textit{independent-particle approximation} (IPA), when the electron-hole interaction is totally neglected.
 
 The square modulus of the transition coefficients
  \begin{equation}\label{eq5_15}
  	t^{\lambda}_{i}=\sum_{cu\mathbf{k}}A^{\lambda}_{cu\mathbf{k}}\frac{\langle c\mathbf{k}|\hat{p}_{i}|u\mathbf{k}\rangle}{\epsilon_{u \mathbf{k}}-\epsilon_{c}}
  \end{equation}
 gives the oscillator strength of the excitation. Both eigenvalues and eigenvectors of Eq. \eqref{eq5_18} enter the expression of the macroscopic dielectric function
 \begin{equation}
   \epsilon^{ij}_{M}(\omega)=\delta_{ij}+\frac{4\pi}{V}\sum_{\lambda}\frac{t^{\lambda}_{i}\left[t^{\lambda}_{j}\right]^*}{\omega-E^{\lambda}+\textrm{i}\delta}.
 \end{equation}
The eigenvectors $A^{\lambda}_{cu\mathbf{k}}$ determine the two-particle wavefunction $\Phi^{\lambda}$, corresponding to an excitation with energy $E^{\lambda}$:
 \begin{equation}\label{eq5_14}
 \begin{aligned}
 	&\Phi^{\lambda}(\mathbf{r}_e s_e, \mathbf{r}_h s_h)=\sum_{cu\mathbf{k}}A^{\lambda}_{cu\mathbf{k}}\psi_{u\mathbf{k}}(\mathbf{r}_e s_e)\psi^*_{c\mathbf{k}}(\mathbf{r}_h s_h)\\
 	&=\sum_{m,m'}\sum_{cu \mathbf{k}} A^{\lambda}_{cu\mathbf{k}} \varphi_{u \mathbf{k}}^{m}(r_{e})\varphi_{c \mathbf{k}}^{m'}(r_{h})^* \eta_{m}(s_e)\eta^*_{m'}(s_h),
\end{aligned} 
 \end{equation}
using the expression $\psi_{n\mathbf{k}}(\mathbf{r}s)= \sum_{m={\uparrow, \downarrow}} \varphi^{m}_{n\mathbf{k}}(\mathbf{r})\eta_{m}(s)$ for the single-particle states in the case of non-collinear spin, with $\eta_m(s)$ being the spinor wavefunction, and $\varphi_{n\mathbf{k}}^{m}$ the spatial one \cite{Roedl}.
 In reciprocal space, the character and the distribution of the $e$-$h$ wavefunction can be expressed by the \textit{weights} of a transition at each $\mathbf{k}$-point:
 \begin{equation}\label{eq5_29}
   w^{\lambda}_{u\mathbf{k}}=\sum_{c}|A^{\lambda}_{cu\mathbf{k}}|^2.
 \end{equation}
 The effective Hamiltonian in the Tamm-Dancoff approximation \bibnote{In the Tamm-Dancoff approximation the coupling between two core states or two conduction states is neglected.
 While it was demonstrated \cite{TDA:Puschnig,TDA:Gruening} that in the optical limit this assumption is justified if the binding energies of the excitations are small compared to the band gap, for calculations at finite momentum transfer the approximation breaks down even if the binding energies are small \cite{TDA:Kresse}. Since momentum transfer is neglected in this work, we do not expect the Tamm-Dancoff approximation to significantly alter our results.} can be separated into three contributions, $H^{BSE}=H^{diag}+H^d+H^x$, where the diagonal term $H^{diag}_{cu\mathbf{k},c'u'\mathbf{k}'}=(\epsilon_{u\mathbf{k}}-\epsilon_{c}+\Delta)\delta_{cc'}\delta_{uu'}\delta_{\mathbf{k}\mathbf{k}'}$ describes transitions within the IPA.
 $H^{x}$ is the repulsive exchange part of the Coulomb kernel
 \begin{eqnarray}\label{eq5_21}
 	H^{x}_{cu\mathbf{k}, c'u'\mathbf{k}'}=&&\int d^3rd^3r' \sum_{m} \varphi_{c\mathbf{k}}^{m}(\mathbf{r})\Big(\varphi^{m}_{u\mathbf{k}m}(\mathbf{r})\Big)^*\bar{v}(\mathbf{r},\mathbf{r}')\nonumber\\&&
   \times\sum_{m'}\left(\varphi^{m'}_{c'\mathbf{k}'m'}(\mathbf{r}')\right)^*\varphi^{m'}_{u'\mathbf{k}'m'}(\mathbf{r}'),
 \end{eqnarray}
 with $\bar{v}(\mathbf{r},\mathbf{r}')$ being the short-range bare Coulomb potential.
 $H^d$ is the attractive direct term
 \begin{eqnarray}\label{eq5_22}
 	H^d_{cu\mathbf{k}, c'u'\mathbf{k}'}=&&-\int d^3rd^3r'\sum_{m}\varphi_{c\mathbf{k}}^m(\mathbf{r})\Big(\varphi^{m}_{c'\mathbf{k}'}(\mathbf{r})\Big)^*W(\mathbf{r},\mathbf{r}')\nonumber\\&&\times\sum_{m'}\Big(\varphi^{m'}_{u\mathbf{k}}(\mathbf{r}')\Big)^*\varphi^{m'}_{u'\mathbf{k}'}(\mathbf{r}'),
 \end{eqnarray}
 where $W(\mathbf{r},\mathbf{r}')$ is the statically screened Coulomb interaction. We note that due to the summations over the quantum number $m=\uparrow, \downarrow$ of the spin projection on the z-component, Eqs. \eqref{eq5_21} and \eqref{eq5_22} are valid for the general case of non-collinear spin polarization.
 \subsection{ All-electron implementation}

 Employing the BSE approach to core spectroscopy requires the explicit determination of the matrix elements between two core and two conduction states involved in the excitation.
To avoid the computationally cumbersome integration over non-local quantities, the matrix elements of Eqs. \eqref{eq5_21}-\eqref{eq5_22} are expanded into a planewave basis.
 In this way, the Fourier-transforms $W_{\mathbf{G}\mathbf{G}'}(\mathbf{q})$ and $\bar{v}_{\mathbf{G}}(\mathbf{q})$ of the screened and bare Coulomb potential, respectively, can be straightforwardly evaluated.
 Using plane wave matrix elements
 \begin{equation}\label{eq5_25}
   M_{ij\mathbf{k}}(\mathbf{q}+\mathbf{G})=\langle \psi_{i\mathbf{k}}|\textrm{e}^{-i(\mathbf{q}+\mathbf{G})}|
   \psi_{j(\mathbf{k}+\mathbf{q})}\rangle,
 \end{equation}
 where $i,j \in [c,u]$, the exchange interaction (Eq.~\ref{eq5_21}) can be written as
 \begin{equation}\label{eq5_26}
   H^{x}_{cu\mathbf{k}, c'u'\mathbf{k}'}=\frac{1}{V}\sum_{\mathbf{G}}M^{*}_{cu\mathbf{k}}(\mathbf{G})\bar{v}_{\mathbf{G}}(\mathbf{q}=0)M_{c'u'\mathbf{k}'}(\mathbf{G}),
 \end{equation}
 and the direct term (Eq.~\ref{eq5_22}) as
 \begin{equation}\label{eq5_26b}
   	H^{d}_{cu\mathbf{k}, c'u'\mathbf{k}'}=-\frac{1}{V}\sum_{\mathbf{G}\mathbf{G}'}M^{*}_{cc'\mathbf{k}}(\mathbf{q}+\mathbf{G})W_{\mathbf{G}\mathbf{G}'}(\mathbf{q})M^{*}_{uu'\mathbf{k}}(\mathbf{q}+\mathbf{G}'),
 \end{equation}
with $\mathbf{q}=\mathbf{k}-\mathbf{k}'$, and $V$ being the unit cell volume.
While, in principle, the summations in both equations include all reciprocal lattice vectors $\mathbf{G}$, in practice they are limited to vectors fulfilling the condition $|\mathbf{G}+\mathbf{q}|\leq |\mathbf{G}+\mathbf{q}|_{max}$.
 The matrix elements of the screened Coulomb potential $W_{\mathbf{G}\mathbf{G}'}(\mathbf{q})$ are obtained as
 \begin{equation}\label{eq5_27}
   W_{\mathbf{G}\mathbf{G}'}(\mathbf{q})=\frac{4\pi \epsilon^{-1}_{\mathbf{G}\mathbf{G}'}}{|\mathbf{q}+\mathbf{G}||\mathbf{q}+\mathbf{G}'|},
 \end{equation}
 where the inverse microscopic dielectric tensor $\epsilon^{-1}_{\mathbf{G}\mathbf{G}'}$ is computed within the random-phase approximation (RPA).
While the long-range term of the potential is only accounted for in the attractive part of the electron-hole interaction of Eq.~\eqref{eq5_26b}, the short-range term appears in both the attractive direct as well as the repulsive exchange interaction. Specifically, the latter describes the so-called \textit{local-field effects} (LFE) \cite{LFE:Adler, LFE:Wiser}.

 Single-particle wavefunctions are obtained from all-electron full-potential DFT calculations, as implemented in the \texttt{exciting} code \cite{exciting}. In this package,  conduction states are computed employing the linearized augmented planewave plus local-orbitals [(L)APW+LO] basis set.
 In this formalism, the electronic wavefunctions are expanded into an atomic part around the nuclear positions defined by the \textit{muffin-tin} (MT) radius $R_{MT}$, whereas in the space between the non-overlapping MT spheres a planewave expansion is adopted. More details about the LAPW+LO method and its implementation in \texttt{exciting} can be found in Ref.~[\onlinecite{exciting}]. Conduction states $\psi^{i\mathbf{k}\alpha}_{\textrm{scalar}}$ are obtained as eigenstates of the scalar-relativistic Hamiltonian. In a MT sphere around the atom $\alpha$, they are expressed most generally as an expansion in spherical harmonics $Y_{lm}(\hat{\mathbf{r}}_{\alpha})$:
 \begin{equation}
   \psi^{i\mathbf{k}\alpha}_{\textrm{scalar}}(\mathbf{r}_{\alpha})=\sum_{lm} u^{i\mathbf{k}}_{l\alpha}(r_{\alpha})Y_{lm}(\hat{\mathbf{r}}_{\alpha}),
 \end{equation}
 where $\mathbf{r}_{\alpha}=\mathbf{r}-\mathbf{R}_{\alpha}$ and $\mathbf{R}_{\alpha}$ is the position of atom $\alpha$.
 To calculate matrix elements between spinorial core states and conduction bands, the latter are assumed to take the form
 \begin{equation}\label{eq5_8}
  	\psi^{i\mathbf{k}}(\mathbf{r} s)=\sum_{m}\frac{1}{\sqrt{2}} \psi_{\textrm{scalar}}^{i\mathbf{k}}(\mathbf{r})\eta_{m}(s),
 \end{equation}
where $\psi_{\textrm{scalar}}^{i\mathbf{k}}$ denotes the eigenstates of the scalar-relativistic Hamiltonian, representing an average between the two spin-channels.
This approximation significantly reduces the computational effort, introducing only a small exchange-like error in the excitonic energies compared to a fully relativistic treatment of the conduction states.

 The initial core states, which are highly localized and dominated by relativistic effects, are obtained as solutions of the Dirac equation in a spherically symmetric potential.
 The full Dirac equation can be transformed into a set of coupled radial equations (for a full derivation, see Refs. [\onlinecite{radialDirac}] and [\onlinecite{radialDirac3}]).
 As the Dirac Hamiltonian commutes with the total angular momentum operator $\hat{\mathbf{J}}=\hat{\mathbf{L}}+\hat{\mathbf{S}}$, it can be decomposed into a radial and a spherical part.
 As such, the four-component Dirac wavefunction is written as \cite{radialDirac2}:
 \begin{equation}\label{eq5_9}
 \psi_{\kappa,M}(\mathbf{r})=\left( \begin{array}{c}u_{\kappa}(r)\Omega_{\kappa,M}(\hat{\mathbf{r}}) \\ -\mathrm{i} v_{\kappa}(r)\Omega_{-\kappa,M} (\hat{\mathbf{r}}) \end{array}\right),
 \end{equation}
 where the quantum number $\kappa$ is introduced as unique index for a state defined by the quantum numbers $^{2S+1}L_{J}$:
 \begin{equation}\label{eq5_11}
  \kappa=\begin{cases} -L-1 & \mbox{for } J=L+\frac{1}{2} \\ \phantom{-}L & \mbox{for } J=L-\frac{1}{2} \end{cases}.
 \end{equation}
 The so-called \textit{spin spherical harmonics} $\Omega_{L,S,J,M}(\hat{r})$ are given by
 \begin{equation}\label{eq5_10}
  \Omega_{(L,\frac{1}{2})L+\frac{1}{2},M}(\hat{r})=\left(\begin{array}{c}\sqrt{\frac{L+M+\frac{1}{2}}
  {2L+1}}Y_{L,M-\frac{1}{2}}(\hat{r}) \\ \sqrt{\frac{L-M+\frac{1}{2}}{2L+1}}Y_{L,M+\frac{1}{2}}(\hat{r}) \end{array}
  \right),
 \end{equation}
 and
 \begin{displaymath}
  \Omega_{(L,\frac{1}{2})L-\frac{1}{2},M}(\hat{r})=\left(\begin{array}{c}-\sqrt{\frac{L-M+\frac{1}{2}}
  {2L+1}}Y_{L,M-\frac{1}{2}}(\hat{r}) \\ \sqrt{\frac{L+M+\frac{1}{2}}{2L+1}}Y_{L,M+\frac{1}{2}}(\hat{r}) \end{array}
  \right).
 \end{displaymath}
In this way, the wavefunctions $\psi_{\kappa,M}$ in Eq.~\eqref{eq5_9} are four-dimensional Dirac vectors, composed of two two-dimensional spinors $\Omega_{\kappa,M}$, where the radial function of the \textit{large} component is $u_{\kappa}(r)$, and that of the \textit{small} component is $-\mathrm{i}v_{\kappa}(r)$.
 The radial functions are obtained by solving the coupled radial Dirac equations \cite{radialDirac}
 \begin{align}
   \begin{split}
  \frac{\partial u_{\kappa}}{\partial r}&= \frac{1}{c}(v_{eff}^{s}-\epsilon_{\kappa})v_{\kappa}+\left( \frac{\kappa-1}{r}\right)u_{\kappa}\\
  \frac{\partial v_{\kappa}}{\partial r}&= -\frac{\kappa+1}{r}v_{\kappa}+2c\left(1+\frac{1}{2c^2}(\epsilon_{\kappa}-v^s_{eff}) \right).
 \end{split}
 \end{align}
 In order to calculate matrix elements with the scalar-relativistic states in the conduction band,  the \textit{small} component of the core states is neglected, thereby producing spinor solutions $\psi_{\kappa,M}(\mathbf{r})$ for a given atom $\alpha$ at position $\mathbf{R}_{\alpha}$.
 This approximation, commonly adopted in full-potential codes (see, \textit{e.g.}, Ref. [\onlinecite{betzinger}]), is required, since the conduction states are solutions of the
 scalar-relativistic Hamiltonian and therefore the \textit{small} component of these states is not accessible.
 The core wavefunction of an atom $\alpha$ finally takes the form
 \begin{equation}\label{eq5_12}
    \psi_{\kappa,M\alpha}(\mathbf{r})=\begin{cases}u_{\kappa\alpha}(r_{\alpha})\Omega_{\kappa,M}(\hat{\mathbf{r}}_{\alpha}) & \mbox{for } r_{\alpha}\leq R_{MT}\\ 0 & \mbox{else }\end{cases}.
 \end{equation}
 Bloch states $\psi_{\kappa,M\alpha}^{\mathbf{k}}$ are obtained from the localized core states as
 \begin{equation}
   \psi_{\kappa,M\alpha}^{\mathbf{k}}(\mathbf{r})=\sum_{\mathbf{R}}\psi_{\kappa,M\alpha}(\mathbf{r}_{\alpha}-\mathbf{R})\textrm{e}^{i\mathbf{k}\mathbf{R}},
 \end{equation}
 where $\mathbf{R}$ is a lattice vector.

\section{Results}
\subsection{Titanium $K$-edge spectra of $\textrm{TiO}_2$}

\begin{figure}[t]
\includegraphics[width=0.85\linewidth]{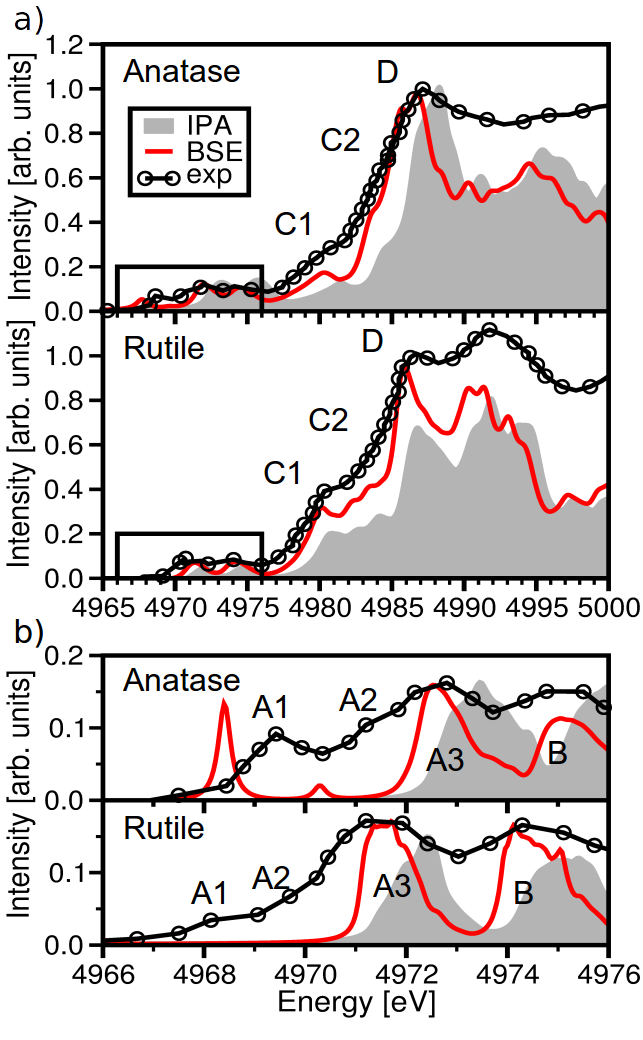}
\caption{\label{fig1:TiO2K} a) Ti $K$-edge spectra of anatase (upper panel) and rutile (lower panel) $\textrm{TiO}_2$. The results obtained  from BSE (red line) are compared with their counterpart from the independent-particle approximation (IPA, grey area), as well as with experimental results (black circles, from Ref.~[\onlinecite{TiO2K_exp}]). A Lorentzian broadening of 0.5 eV is used for both theoretical spectra. b) Pre-edge features of rutile (lower panel) and anatase (upper panel). A smaller Lorentzian broadening of 0.1 eV is adopted to resolve individual features. The BSE absorption spectra are obtained from an average over the diagonal components of the dielectric tensor and are aligned to the experimental ones at the peak A3. The labels denote the peaks in the experimental spectra. }
\end{figure}

We start our analysis by investigating near-edge excitations from the Ti $1s$ electrons in the rutile and anatase phases of $\textrm{TiO}_2$. Due to their tetragonal crystal symmetry, these materials are characterized by two inequivalent optical components, corresponding to parallel and perpendicular light polarization. The computed x-ray near-edge absorption spectra (XANES) shown in Fig.~\ref{fig1:TiO2K}a represent an average of both, as no information about the experimental setup is provided \cite{TiO2K_exp}. We choose to align the computed spectrum with respect to the peak A3, since the latter appears with the same oscillator strength in all optical components.
Excitations from the Ti $K$ edge occur at very high energy, \textit{i.e.}, approximately 5 keV, and are therefore probed by hard x-rays.
Three distinct peaks are visible in the spectra.
Following the notation adopted in Ref. \cite{TiO2K_exp}, C1 and C2 are the so-called edge-crest peaks, while D is the first edge peak.
In the rutile phase, C1 has relatively large oscillator strength, whereas C2 is barely visible as a shoulder of D.
Our results reproduce very well the experimental data of both $\textrm{TiO}_2$ phases in terms of position and relative intensity of the peak D.
The comparison with the IPA spectrum, also shown in Fig.~\ref{fig1:TiO2K}a, reveals that at the absorption onset excitonic effects do not play a major role.
Except for a natural blue-shift of the peaks due to the missing electron-hole interaction, the overall spectral shape, which is very similar in both phases, is qualitatively captured also by a mean-field description of the transitions.
This is not surprising, considering that the latter involve unoccupied states in the continuum.

 \begin{figure}
 \includegraphics[width=0.95\columnwidth]{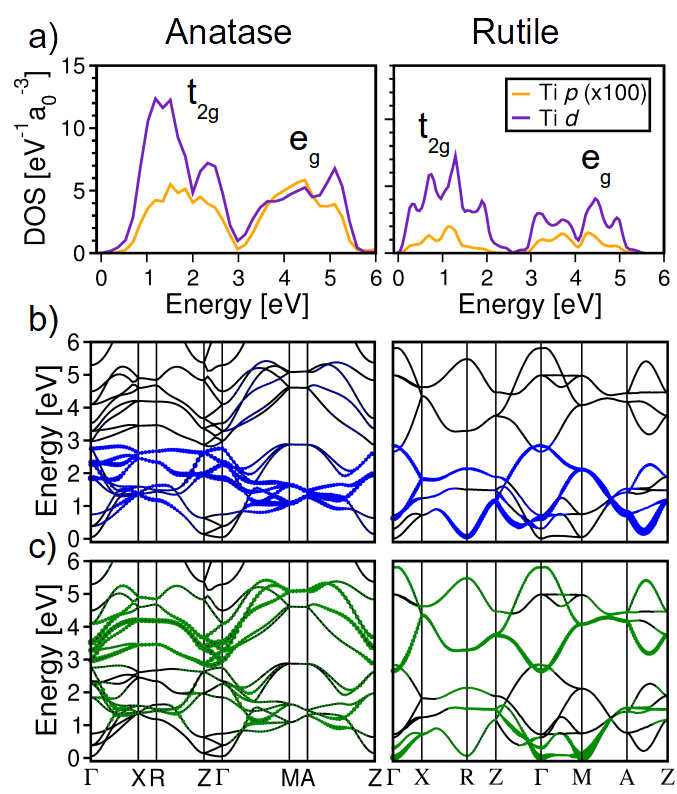}%
 \caption{\label{fig2:TiO2K_exciton}a) Projected density of states (DOS) in the conduction-band region of rutile and anatase. The Ti $p$ contributions are multiplied by a factor 100 to enable their visualization. Contributions of individual bands to the excitons in the b) A1 and c) A2 pre-peaks in the Ti $K$ edge spectra of anatase (left), and rutile (right). The size of the colored circles, blue for A1 (top) and green for A2 (bottom), quantifies the weight of each band in the excitonic eigenstates. Energies are given relative to the onset of the conduction band.}
 \end{figure}
Correlation effects come relevantly into play in the pre-edge features, which also determine the XANES fingerprints of the two phases (Fig.~\ref{fig1:TiO2K}b), as extensively discussed in Refs. \cite{TiO2K_theo1,TiO2K_theo2,TiO2K_theo3,TiO2K_theo4,TiO2K_BSE,TiO2K_BSE2}.
In anatase the lowest-energy feature corresponds to an intense bound exciton, A1, with a binding energy of 3.42 eV.
About 2 eV higher in energy, a weaker peak, A2, appears with a binding energy of 1.54 eV. These two peaks show a strong polarization dependence: While A1 appears only in the optical component perpendicular to the tetragonal axis, A2 is only present in the parallel component.
On the contrary, in rutile these two excitations are dark and the first active peak is A3.
Remarkably, a similar behavior occurs also in the optical region, where the anatase phase exhibits bright bound excitons, albeit at these frequencies binding energies are two orders of magnitude smaller than in the core spectrum \cite{chio+10prb,kang-hybe10prb}.
The high-energy part of the pre-edge spectra is characterized by two intense and broad peaks labeled A3 and B, stemming from transitions into the low-energy part of the conduction region.
As shown in Fig.~\ref{fig2:TiO2K_exciton}a, the first unoccupied states in both phases are dominated by Ti $3d$ states hybridized with the $4p$ ones.
The contribution from the latter is crucial to make excitations into these bands dipole-allowed.
Obviously, the low oscillator strength of the pre-edge features is consistent with the dominant $3d$ character of the lowest-energy conduction bands.

The distinct pre-edge spectral signatures of rutile and anatase can be ascribed to the differences in the unoccupied electronic structure of the two phases, and, in turn, to the Coulomb interaction between the core-hole and the excited electron.
In Fig.~\ref{fig2:TiO2K_exciton}b-c we show the contributions to the excitations from different conduction states, by projecting the coefficients $w^{\lambda}_{c\mathbf{k}}$ (Eq. \ref{eq5_29}) onto the respective band
structure.
In both phases, the lowest-energy exciton A1 stems from transitions to the $t_{2g}$ band (see also Fig.~\ref{fig2:TiO2K_exciton}a) with predominantly Ti $3d$ character.
The distribution of the exciton weights over the entire Brillouin zone points to a localization in real space, typical of bound excitons.
It is worth noting that the contributions coming from all the bands of the $t_{2g}$ manifold are an indication of the importance of correlation effects.
A closer look at the projected density of states in the conduction region (Fig.~\ref{fig2:TiO2K_exciton}a) reveals that in anatase the hybridization of the Ti $3d$ band with $4p$ states in this energy window is enhanced compared to rutile.
Since only transitions to Ti $4p$ states are allowed by the dipole selection rules, the oscillator strength of A1 is non-zero in the former phase, while vanishing in the latter.
The second exciton, A2, involves in both phases also transitions to the $e_g$ manifold, at higher energies.
Again, the relative weight of $4p$ states is larger in anatase compared to rutile, and therefore the intensity of A2, although very weak, is non zero.
Also in this case, the analysis of the exciton weights reveals that the electron-hole correlation induces a remarkable mixing of single-particle transitions over the whole Brillouin zone.

Previous results obtained from multiple-scattering calculations \cite{TiO2K_theo1,TiO2K_theo2,TiO2K_theo3,TiO2K_theo4} correctly describe the overall spectral shape in both phases, but show limitations regarding the bound excitons in the pre-edge region.
A better description of the latter features is achieved by a BSE treatment \cite{TiO2K_BSE,TiO2K_BSE2}, and further improved by the all-electron approach adopted in this work. Overall, we find good agreement with experimental data \cite{TiO2K_exp,TiO2K_exp2}, considering that the latter are a result of indirect measurements and, as such, may include scattering, losses and higher-order effects.
We notice, though, that in anatase the intensity of the first peak A1 is overestimated in our calculation, while the second one, A2, is too weak.
The absolute energy of these two excitations is red-shifted by approximately 1 eV compared to the experiment.
This discrepancy is likely due to the adopted spinor approximation for the conduction states (Eq.~\ref{eq5_8}). In this way, the repulsive exchange interaction is slightly underestimated, resulting in the observed red-shift of the bound excitons.
Quasi-particle corrections to the energy eigenvalues of the correlated Ti $3d$ conduction bands may further improve the relative peak positions. Additionally, lattice contributions to the screening, which are neglected in the current work, might further influence the binding energies of the bound excitons in the pre-edge region.

\subsection{Lead $M_{4}$-edge spectrum of $\textrm{PbI}_2$}

\begin{figure}
\includegraphics[width=0.95\columnwidth]{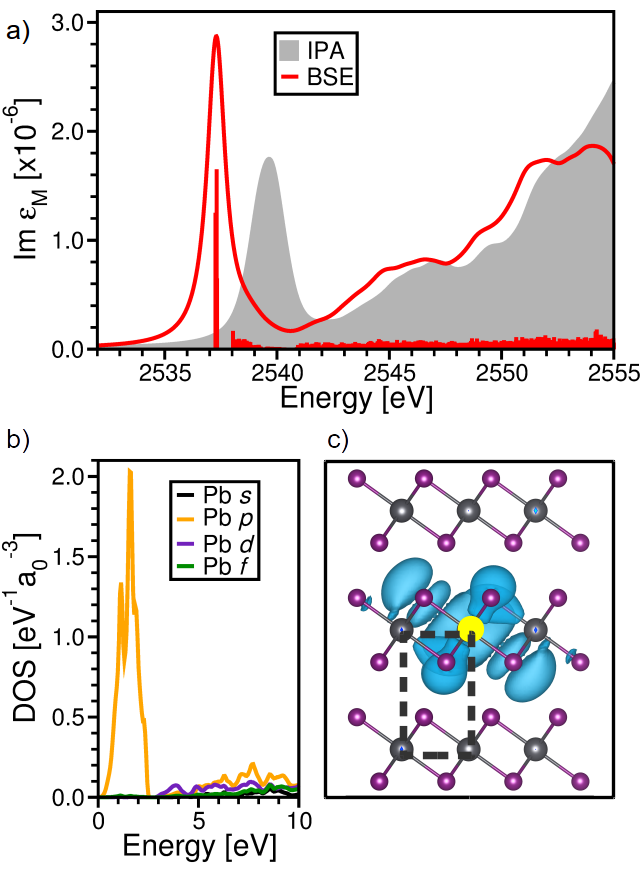}
\caption{\label{fig4:PbI2M}a) Pb $M_4$-edge absorption spectrum of lead iodide obtained from BSE (red line) and within the independent-particle approximation (IPA, shaded gray area). The oscillator strength of the BSE excitations is indicated by vertical bars. A Lorentzian broadening of 0.5 eV is used in both calculated spectra to mimic the excitation lifetime. The calculated absorption spectrum is obtained from an average over the diagonal components of the dielectric tensor. b) Density of states projected onto the Pb $s$, $p$, $d$, and $f$ states in $\textrm{PbI}_{2}$. The onset of the conduction band is set to zero. c) Real-space representation of the electron distribution of the lowest-energy exciton for a fixed position of the core hole. The absorbing atom is marked in yellow. Iodide atoms are displayed in purple, lead atoms in gray.}
\end{figure}

As a second case, we study the XANES of lead iodide from the Pb $M_{4,5}$ edge. The Pb $3d$ states are deep core levels, but their binding energies are only about half of that of the Ti $1s$ states. Although, in principle, transitions from all Pb $3d$ states have to be considered, in practice, the $M_5$ and $M_4$ edges can be treated separately. As the spin-orbit splitting between the Pb $3d_{3/2}$ and the $3d_{5/2}$ electrons is about 104 eV, transitions from these initial states cannot interact significantly.
In the following, we focus only on the $M_4$ edge, since the $M_5$ edge spectrum displays the same features.

The dielectric function shown in Fig.~\ref{fig4:PbI2M}a is characterized by a sharp peak at 2537 eV, followed by two broader features at approximately 2546 eV and 2554 eV, respectively.
The first peak is formed by two bound excitons with large oscillator strength and binding energies of 0.95 eV and 0.90 eV, respectively.
A second group of excitations occurs between 2538 eV and 2540 eV, contributing to the first peak as a weak shoulder.
Above 2541 eV, a continuum of low-intensity excitations characterizes the spectrum, giving rise to a broad hump.
In order to understand the nature of these excitations, we now compare the spectrum computed from the solution of the BSE with the one obtained in the IPA. The spectral shape is essentially the same with and without excitonic effects. However, the electron-hole interaction shifts the spectral weight, thereby increasing the oscillator strength of the excitations at the onset. This is the typical behavior exhibited in the optical region by bound excitons in solid-state materials \cite{BSE:Onida}. However, it should be noted that bound excitons in core spectra typically have much larger binding energies. In the case of $\textrm{PbI}_2$, the binding energy of the first bound exciton is about 3 eV, two orders of magnitude larger than the first optical excitation \cite{PbI2_Theo}.
These indications confirm the dominant role of the screened Coulomb interaction in ruling the core excitations in lead iodide from the Pb $M_4$ edge.

The intense feature at the absorption onset can be analyzed in terms of the projected density of states (Fig.~\ref{fig4:PbI2M}b).
We notice that the low-energy region is dominated by an intense peak coming from the Pb $p$ states, which are only very weakly hybridized with Pb $f$ states.
Transitions to these states are dipole allowed and therefore give rise to the large oscillator strength at the onset of the IPA spectrum. The excitonic nature of the lowest-energy excitation obtained from BSE is revealed by the analysis of the real-space distribution of the electron-hole pair (Fig.~\ref{fig4:PbI2M}c).
While the electron probability is overall localized around the absorbing lead atom,
significant contributions also come from the six nearest-neighbor iodide atoms.
The correlated electron distribution for a fixed core-hole position, which exceeds the unit cell (dashed line in Fig.~\ref{fig4:PbI2M}c), is confined within the layer of the absorbing atom.
Its extension beyond the size of the unit cell of the material is a signature of the electron-hole correlation, which is typical of bound excitons in the core \cite{Ga2O3,Olovsson,Olovsson2, Olovsson3}, as well as in the valence region (see, \textit{e.g.}, \cite{BSE:Rohlfing,PuschnigPaper}).
It is worth noting that the first core exciton in $\textrm{PbI}_2$ is much more localized than its counterpart in the optical region \cite{PbI2_Theo}.

\subsection{Calcium $L_{2,3}$-edge spectrum of CaO}

\begin{figure}
\includegraphics[width=0.85\columnwidth]{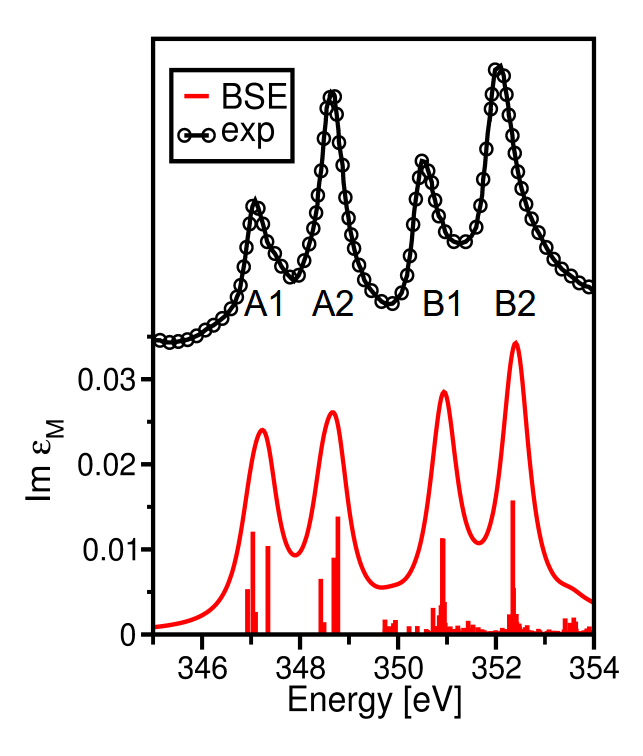}
\caption{\label{fig3:CaOL} Ca $L_{2,3}$-edge absorption spectrum of CaO expressed by the imaginary part of the dielectric function (red line) compared with experimental data (black dots, from Ref. [\onlinecite{CaO_ref}]) offset for clarity. The oscillator strength of individual excitations is indicated by vertical bars. A Lorentzian broadening of 0.3 eV is used in the calculated spectrum to mimic the excitation lifetime.}
\end{figure}
As a final case, we study the excitations from the Ca $2p_{1/2}$ and $2p_{3/2}$ states in CaO. Here, the core levels are considerably higher in energy than the Ti $1s$ and Pb $3d$ states considered before. Moreover, they are separated by a spin-orbit splitting of about 3.7 eV. Thus, transitions from both Ca $2p_{1/2}$ and $2p_{3/2}$ initial states have to be considered simultaneously.
The XANES from the Ca $L_{2,3}$ edge of CaO shown in Fig.~\ref{fig3:CaOL} exhibits the prototypical \textit{white-line} form of $L_{2,3}$-edge spectra also found in many $3d$ transition metals \cite{TM_whitelines}.
The spectral shape is dominated by four intense excitonic peaks, associated with the $L_3$ (A1 and A2, following the notation used in Ref. \onlinecite{CaO_ref}) and the $L_2$ (B1 and B2) sub-edges.
This spectrum is strongly affected by correlation effects due to the considerable $e$-$h$ exchange interaction.
This type of interaction is associated with the local fields acting in the system, generated by the anisotropic character of electronic charge distribution and described by the short-range part of the bare Coulomb interaction (Eq. \ref{eq5_26}).

\begin{figure}[t]
\includegraphics[width=0.95\columnwidth]{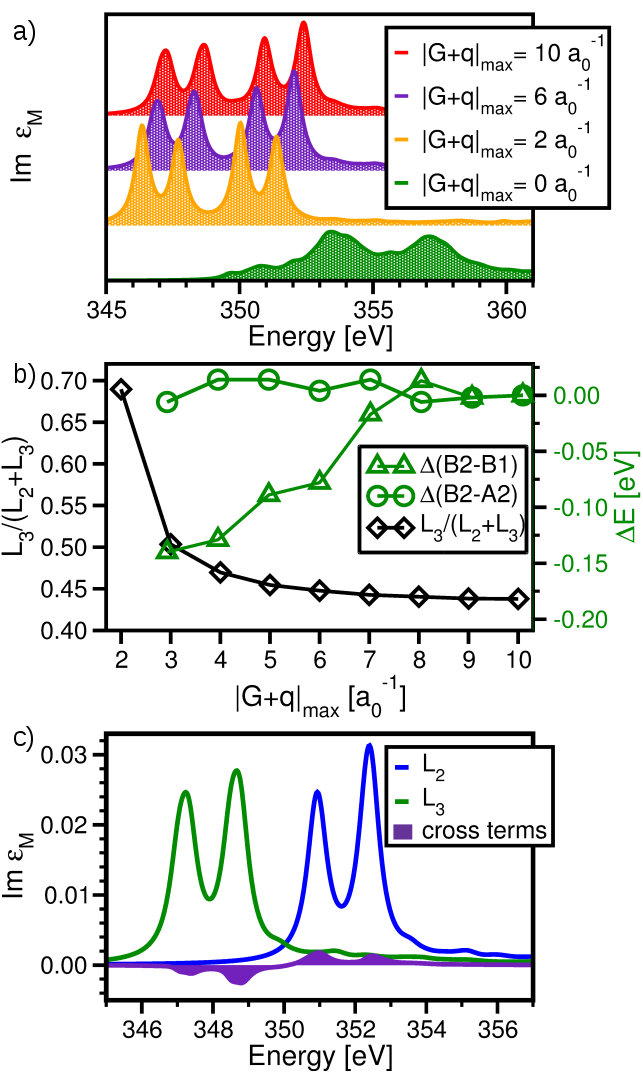}
\caption{\label{fig3:CaOLcross}a) Ca $L_{2,3}$-edge spectra of CaO computed at increasing values of $|\mathbf{G}+\mathbf{q}|_{max}$. b) $|\mathbf{G}+\mathbf{q}|_{max}$ dependence of the energy differences between the peaks B1 and B2, $\Delta (\textrm{B}2-\textrm{A}2)$ (circles), as well as between B2 and A2, $\Delta (\textrm{B}2-\textrm{B}1)$ (triangles), and the branching ratio $L_3/(L_2+L_3)$ (diamonds). The former are relative to the final values of (B2-A2) = 3.69 eV and (B2-B1) = 1.46 eV. c) Absorption spectra of CaO obtained including only Ca $2p_{1/2}$ ($L_2$) and Ca $2p_{3/2}$ ($L_3$) as initial states. The \textit{cross terms}, $L_{2,3}-(L_2+L_3)$, between the total $L_{2,3}$ spectrum and the individual $L_2$ and $L_3$ ones are also shown (solid area).}
\end{figure}

To explain the essential role of LFE in capturing the features of the Ca $L_{2,3}$-edge spectrum of CaO, we consider the result obtained by neglecting the exchange term in the solution of the BSE Hamiltonian.
In this scenario [setting $|\mathbf{G}+\mathbf{q}|_{max}=0$ in Eqs. \eqref{eq5_26} and \eqref{eq5_26b}], the screened Coulomb potential in Eq. \eqref{eq5_27} gives a non-zero contribution, while $\bar v$ vanishes by definition.
As discussed in Refs. \cite{Cocchi,PuschnigPhD, PuschnigPaper} for optical spectra, in this case the screened Coulomb interaction merely induces a rigid red-shift of the IPA spectrum.
Indeed, the resulting spectrum shown in Fig.~\ref{fig3:CaOLcross}a is very broad and rather featureless, with no excitonic peaks appearing.
This is a clear indication that the $\mathbf{G}$=0 term of the $e$-$h$ attraction alone is not capable of describing the core excitations.
However, as indicated in Fig.~\ref{fig3:CaOLcross}a, it is sufficient to take into account only 24 $|\mathbf{G}+\mathbf{q}|$ vectors, corresponding to a cutoff $|\mathbf{G}+\mathbf{q}|_{max}=2 \; a_0^{-1}$, to obtain the correct spectral shape, although with inaccurate relative peak heights.
Thus, the screened Coulomb attraction between electron and core hole is well captured already with a relatively low $|\mathbf{G}+\mathbf{q}|_{max}$ cutoff. Conversely, the repulsive electron-hole exchange, which mixes transitions from the two sub-edges, requires a more accurate description of local-field effects.
It is worth noting that even our best result, obtained by including approximately 3000 $|\mathbf{G}+\mathbf{q}|$ vectors, corresponding to a cutoff of 10 $a_0^{-1}$, does not perfectly reproduce the strength of the measured peaks (Fig.~\ref{fig3:CaOL}), since the relative intensity between A1 (B1) and A2 (B2) is still underestimated.
 \begin{table} 
 \caption{\label{table1}Energy differences B1-B2, B2-A2, and branching ratio $L_3/(L_2+L_3)$ for the Ca $\textrm{L}_{2,3}$-edge spectrum of CaO. The values computed from BSE are compared with available experimental data and with results obtained with the configuration interaction (CI) method.}
 \begin{ruledtabular}
 \begin{tabular}{lccc}
  Method & B1-B2 [eV] & B2-A2 [eV] & $L_3/(L_2+L_3)$ \\
  \hline
  BSE (this work) & 1.46 & 3.69 & 0.44 \\
  CI \cite{CaO_ref} & 1.51 & 3.60 & 0.42 \\
  Experiment \cite{CaO_ref} & 1.53 & 3.40 & 0.38
 \end{tabular}
 \end{ruledtabular}
 \end{table}

The trends described above are summarized in Fig.~\ref{fig3:CaOLcross}b, where the energy differences between the peaks within each sub-edge (triangles) and between the two sub-edges (circles) are plotted at increasing planewave cutoff.
On the one hand, the relative distance between the more intense peaks A2 and B2, given by transitions from the Ca $2p_{3/2}$ and the $2p_{1/2}$ states, respectively, is essentially independent of how accurately LFE are treated, provided that the latter are not completely neglected.
On the other hand, the energy separation between the two peaks, B1 and B2, within the $L_2$ edge is very sensitive to LFE: Only cutoff values larger than 9 $a_0^{-1}$ ensure a reliable result.
The comparison with literature data reported in Table~\ref{table1} corroborates this conclusion.

The relative intensity of the peaks between the two sub-edges also depends crucially on the treatment of LFE.
This value, commonly called \textit{branching ratio} and indicated in Fig.~\ref{fig3:CaOLcross}b as $L_3/(L_2+L_3)$, is expected to be equal to 2/3 (the so-called \textit{statistical value}) in a pure single-particle picture.
The statistical branching ratio arises from the fact that $2p_{1/2}$ states are two-fold degenerate while the $2p_{3/2}$ ones are four-fold degenerate.
Hence, ignoring spin-orbit coupling, the intensity of the $L_3$ sub-egde is expected to be twice as large as the $L_2$ one.
However, in the XANES from the $L_{2/3}$ edge of metal oxides the branching ratio is typically smaller than the statistical value.
Indeed, the strong exchange interaction induces a mixing between transitions from the $2p_{1/2}$ and $2p_{3/2}$ initial states \cite{branchingI,Blaha,CaO_ref}.
A similar behavior occurs also for CaO.
To quantify this effect, we consider the integrated oscillator strength of the Ca $L_3$  and $L_2$ sub-edges.
Such analytic integration is performed at increasing values of $|\mathbf{G}+\mathbf{q}|_{max}$ in the energy range of 336 -- 350 eV for the $L_3$ and of 350 -- 357 eV for the $L_2$ edge.
A Lorentzian broadening of 0.5 eV is adopted in all cases.
The resulting branching ratios are plotted in Fig.~\ref{fig3:CaOLcross}b (diamonds).
For the lowest non-zero cutoff value, $L_3/(L_2+L_3)$=0.69, which is very close to the statistical value.
With increasing $|\mathbf{G}+\mathbf{q}|_{max}$, the branching ratio exhibits a monotonic decrease, down to 0.44, obtained for  $|\mathbf{G}+\mathbf{q}|_{max}> 8 \; a_0^{-1}$.
This result indicates that only a very accurate description of the $e$-$h$ correlation in terms of the exchange interaction can yield the correct spectrum.
Our results are in good agreement with available experimental data, as well as with calculations based on the configuration interaction method \cite{CaO_ref} (see Table~\ref{table1}).

The XANES computed for the individual sub-edges can be compared with the full $L_{2,3}$ spectrum, in order to evaluate the role of mixing between the two initial states $2p_{3/2}$ and $2p_{1/2}$.
To do so, we show in Fig.~\ref{fig3:CaOLcross}c also the so-called \textit{cross terms} [$L_{2,3}-(L_{2}+L_{3})$]. This plot indicates that indeed part of the intensity coming from the $L_3$ sub-edge is transferred to the $L_2$ one.
In CaO such an effect is not as strong as in transition-metal oxides \cite{Blaha}, but still relevant for an accurate description of the spectra.

\section{Conclusions}

We have presented an in-depth analysis of electron-hole correlation effects in core excitations, by studying the XANES of three prototypical solid-state materials, namely $\textrm{TiO}_2$ (rutile and anatase phases), $\textrm{PbI}_2$, and CaO, considering excitations from the $K$, the $M_4$, and the $L_{2,3}$ absorption edges, respectively.
By means of an accurate \textit{ab initio} approach based on the solution of the Bethe-Salpeter equation in an all-electron framework, we have shown that electron-hole correlation is always crucial in the description of these spectra, although manifesting itself in different forms.
For transitions from very deep core levels ($>$1 keV), such as from the Ti $1s$ and the Pb $3d_{3/2}$ states, the attractive $e$-$h$ interaction dominates the absorption process, giving rise to localized bound excitons.
On the other hand, for excitations from the Ca $L_{2,3}$ edge in the metal-oxide CaO, occurring at around $\sim$350 eV, an accurate treatment of the electron-hole exchange becomes crucial.
For this purpose, a careful description of local-field effects turns out to be essential to correctly reproduce the relative intensity of the peaks, as well as the energy difference between them.
The correct spin-orbit splitting between the initial states is captured through an explicit description of core electrons, as enabled by the all-electron formalism adopted in this work. The inclusion of momentum transfer and the extension of this formalism to include quasi-particle correction in the underlying electronic structure, including core states, is envisaged to further improve the predictive power of this methodology.

\appendix*
\section{Computational details}
In the study of the Ti $K$-edge absorption spectra of $\textrm{TiO}_2$, we adopt the lattice parameters $a=b=4.64$ \AA{} and $c=2.97$ \AA{} for rutile (simple tetragonal structure, space group $P42/mnm$), as well as $a=b=3.73$ \AA{} and $c=9.76$ \AA{} for anatase (body-centered tetragonal structure, space group $I4_1/amd$).
These values, in good agreement with experimental data \cite{TiO2_struct_rutile, TiO2_struct_anatase}, are obtained from a full lattice relaxation \cite{OlgaMaster}.
Ground-state calculations are performed using a \textbf{k}-grid of $4\times4\times6$, and a planewave cut-off $R_{MT}|\mathbf{G}+\mathbf{k}|_{max}=12$ for rutile, and a \textbf{k}-grid of $6\times6\times2$, and $R_{MT}|\mathbf{G}+\mathbf{k}|_{max}=7$ for anatase. MT spheres with radius $R_{MT}=1.8 \; a_0$ were employed for both elements in both phases.
For the calculation of the core absorption spectrum of rutile we adopt $R_{MT}|\mathbf{G}+\mathbf{k}|_{max}= 7$ and   $6\times 6\times 9$ $\mathbf{q}$-grid shifted by $\Delta \mathbf{q}=(0.05, 0.15, 0.25)$.
To treat LFE we adopt a cut-off of $|\mathbf{G}+\mathbf{q}|_{max}=4 \; a_0^{-1}$.
The screening of the Coulomb potential is calculated in the random-phase approximation (RPA), including all valence bands and 100 unoccupied ones.
50 unoccupied bands are considered in the diagonalization of the BSE Hamiltonian.
For anatase, the spectrum is obtained with $R_{MT}|\mathbf{G}+\mathbf{k}|_{max}= 7$ and a shifted $8\times 8\times 3$ $\mathbf{q}$-grid  [$\Delta \mathbf{q}=(0.05, 0.15, 0.25)$], with a planewave cut-off of $|\mathbf{G}+\mathbf{q}|_{max}=3 \, a_0^{-1}$ to account for LFE.
70 unoccupied states are included in the diagonalization of the BSE Hamiltonian.
Scissors operators of 113 eV (rutile) and 114 eV(anatase) are applied to align the peak A3 of the BSE spectra with its experimental counterpart, extracted from Ref.~[\onlinecite{TiO2K_exp}].
The same scissors shifts are adopted for the corresponding IPA spectra.

We investigate $\textrm{PbI}_2$ in the hexagonal phase (space group $P\bar{3}m1$).
Lattice parameters of $a=4.54$ \AA{} and $c=6.98$ \AA{}, obtained from volume optimization, are in good agreement with the experimental values \cite{PbI2:structure}.
Ground-state calculations are performed using a $6\times6\times4$ \textbf{k}-mesh, a planewave cut-off $R_{MT}|\mathbf{G}+\mathbf{k}|_{max}=10$, and MT spheres with radius $R_{MT}=2 \; a_0$ for both elements.
The absorption spectrum from the Pb $\textrm{M}_{4}$ edge is calculated with $|\mathbf{G}+\mathbf{q}|_{max}=2$, a shifted $8\times 8\times 6$ $\mathbf{q}$-grid, $R_{MT}|\mathbf{G}+\mathbf{k}|_{max}=10$, 30 conduction bands in the BSE Hamiltonian, and 100 unoccupied ones in the RPA screening. The $\mathbf{q}$-shift is identical to the one used for $\textrm{TiO}_2$. We use a muffin-tin radius $R_{MT}=2.6 \; a_0$ for both elements.
In absence of any published experimental reference, we have not applied any scissors operator to our calculated spectrum.
Thus, we expect the absorption onset to be underestimated by approximately 50 eV compared to the experimental one.
For the visualization of the excitonic wavefunction, the core hole is shifted with respect to the position of the absorbing Pb atom by 7\% along the $x$-axis, to avoid the node of the $3d$ wavefunctions.
The corresponding plot is produced with the VESTA software \cite{VESTA}.

CaO is a cubic material (space group $Fm\bar{3}m$, number 225).
Volume optimization yields a lattice constant of
$a=4.76$ \AA{} in good agreement with the experimental one 
\cite{CaO_struct}.
Ground-state calculations are performed using a \textbf{k}-mesh of $7\times7\times\times7$,  a planewave cut-off $R_{MT}|\mathbf{G}+\mathbf{k}|_{max}=12$, and MT spheres with radius $R_{MT}=2\; a_0$ for both elements.
The $L_{2,3}$-edge spectrum is calculated using a shifted $5\times 5\times 5$ $\mathbf{q}$-grid [$\Delta \mathbf{q}=(0.05, 0.15, 0.25)$], $R_{MT}|\mathbf{G}+\mathbf{k}|_{max}= 7.9$, 20 conduction bands in the BSE Hamiltonian, and 100 unoccupied bands in the RPA calculation of the screening.
A planewave cut-off $|\mathbf{G}+\mathbf{q}|_{max}=10 \; a_0^{-1}$ is adopted to account for LFE.
A scissors operator of 23.5 eV is applied to align the BSE spectrum to the peak A2 in the experimental one, extracted from Ref.~[\onlinecite{CaO_ref}].
The same shift is applied also to the IPA spectrum. The intensity plot in Fig.~\ref{fig1:TiO2K} is normalized to $6\times 10^{-6}$ for rutile, and to $2.5\times 10^{-6}$ for anatase.

The above-listed computational parameters adopted for the BSE spectra ensure a reliable convergence of the spectral shape, as well as an accuracy of 10 meV for the lowest-energy eigenvalue for all investigated systems.

Input and relevant output files of the calculations can be found on the NoMaD Repository, the corresponding DOI is shown in Ref. \bibnote{\url{http://dx.doi.org/10.17172/NOMAD/2016.12.06-1}}

\begin{acknowledgments}
This work was partly funded by the German Research Foundation (DFG), through the Collaborative Research Centers SFB 658 and 951.
C.V. acknowledges financial support from the Humboldt Research Track Scholarship of the Humboldt-Universit\"at Berlin.
C.C. acknowledges funding from the \textit{Berliner Chancengleichheitsprogramm} (BCP) and from IRIS Adlershof.
\end{acknowledgments}

\end{document}